\documentclass[journal=nalefd,manuscript=article, layout=onecolumn]{achemso}

\usepackage{chemformula} 
\usepackage[T1]{fontenc} 
\usepackage{tabularx}
\usepackage{amsmath}
\usepackage[colorinlistoftodos,prependcaption,textsize=footnotesize]{todonotes}
\usepackage[shortcuts]{extdash}

\author{Florian Johst}
\affiliation{Institute of Physical Chemistry, University of Hamburg, Grindelallee 117, D-20416 Hamburg, Germany}
\author{Jannik Rebmann}
\affiliation{Institute of Physical Chemistry, University of Hamburg, Grindelallee 117, D-20416 Hamburg, Germany}
\author{Hans Werners}
\affiliation{Institute of Physical Chemistry, University of Hamburg, Grindelallee 117, D-20416 Hamburg, Germany}
\author{Lars Klemeyer}
\affiliation{Institute of Nanostructure and Solid State Physics, University of Hamburg, Luruper Chaussee 149, D-22761 Hamburg, Germany}
\author{Jagadesh Kopula Kesavan}
\affiliation{Institute of Nanostructure and Solid State Physics, University of Hamburg, Luruper Chaussee 149, D-22761 Hamburg, Germany}
\author{Dorota Koziej}
\affiliation{Institute of Nanostructure and Solid State Physics, University of Hamburg, Luruper Chaussee 149, D-22761 Hamburg, Germany}
\author{Christian Strelow}
\affiliation{Institute of Physical Chemistry, University of Hamburg, Grindelallee 117, D-20416 Hamburg, Germany}
\author{Gabriel Bester}
\affiliation{The Hamburg Centre for Ultrafast Imaging, Luruper Chaussee 149, D-22761 Hamburg, Germany}
\author{Alf Mews}
\affiliation{Institute of Physical Chemistry, University of Hamburg, Grindelallee 117, D-20416 Hamburg, Germany}
\author{Tobias Kipp}
\email{tobias.kipp@uni-hamburg.de}
\affiliation{Institute of Physical Chemistry, University of Hamburg, Grindelallee 117, D-20416 Hamburg, Germany}

\title[An \textsf{achemso} demo]
    {Exciton--Phonon Coupling in Single Band-Gap Engineered ZnCdSe-Dot/CdS-Rod Nanocrystals}
    
\begin{document}
\begin{abstract}
Exciton--phonon coupling limits the homogeneous emission linewidth of nanocrystals. Hence, a full understanding of it is crucial. In this work, we statistically investigate exciton--phonon coupling by performing single-particle spectroscopy on Zn$_{1-x}$Cd$_{x}$Se/CdS dot-in-rod nanocrystals at cryogenic temperatures ($T\approx 10~\rm{K}$). In situ cation exchange enables us to analyze different band alignments and thereby different charge-carrier distributions. We find that the relative intensities of the longitudinal optical S- and Se-type phonon replicas correlate with the charge-carrier distribution. Our experimental findings are complemented with quantum mechanical calculations within the effective mass approximation that hint at the relevance of surface charges.
\end{abstract}

Semiconductor nanocrystals (NCs) are promising building blocks for optoelectronic devices.
\cite{Kershaw.2017,Panfil.2018,Kagan.2019,Harankahage.2021,Efros.2021,Micheel.2022,NobelPrize.2023} Their properties can be fine-tuned by controlling their morphology and composition. Dot-in-rod nanostructures (DRs)\cite{Talapin.2003,Dorfs.2008} are particularly interesting, as they offer intrinsically polarized fluorescence emission \cite{Talapin.2003,Planelles.2016,Vezzoli.2015,Bai.2022} and high extinction coefficients\cite{Talapin.2003}. They can be used as gain media in lasers\cite{Grivas.2013,Manfredi.2018}, in catalysis\cite{OConnor.2012,Wu.2014,Wachtler.2016}, and in biosensing\cite{Kuo.2018,Park.2018}. The properties of NCs are determined by the exciton localization, which has been investigated by time-resolved photoluminescence\cite{Raino.2011,Morgan.2019} or scanning tunneling spectroscopy\cite{Millo.2001}. By its nature, exciton--phonon coupling also contains indirect information about the exciton localization. It is expressed in the form of phonon-replica in fluorescence spectra, even resolvable for single monolayers.\cite{Chilla.2008} However, the experimental methods and theoretical models show discrepancies in the magnitude of exciton--phonon coupling.\cite{Kelley.2019,Lin.2023} Therefore, finding a clear correlation between exciton localization and phonon coupling is essential for understanding the fundamentals of exciton--phonon coupling.

Here, we investigate how the exciton--phonon coupling correlates with the exciton localization in heterostructured DRs. Based on alloyed Zn$_{1-x}$Cd$_{x}$Se/CdS DRs we have developed a system with a gradual change from type\=/II to type\=/I band alignment. By controlling the Cd fraction $x$ the band alignment and the corresponding exciton localization was manipulated, making it a suitable test bed for a gradual change of phonon coupling. The exciton--phonon coupling is investigated from over 300 single particle PL-spectra at cryogenic temperatures. We find that the relative intensities of the S- and Se-type LO phonon replicas reflect the different exciton distributions. The experimental data are compared to a theoretical modeling of exciton--phonon coupling. To this end, we performed quantum mechanical calculations of the excitons within the effective mass approximation, including the Coulomb interaction between electron and hole.

\begin{figure*}[]
    \centering
    \includegraphics[width=13.5 cm]{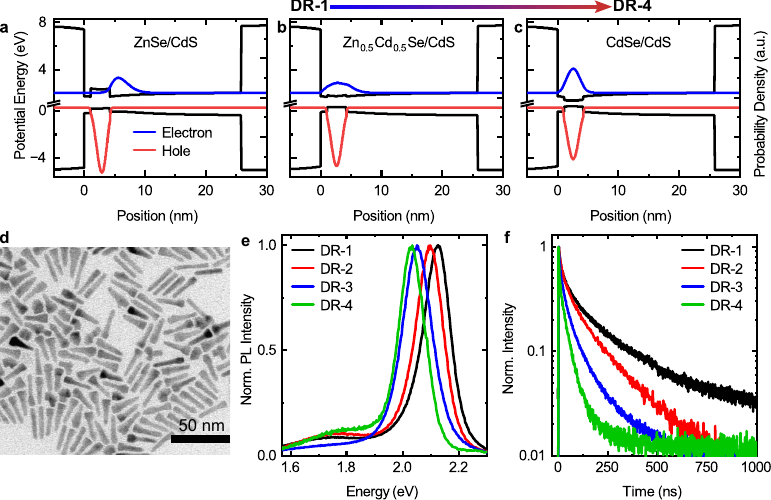}
    \caption{(a-c) Cross sections of the calculated potential energies (black) and probability densities of electron (blue) and hole (red) along the DR axis. (d) TEM image of Zn$_{1-x}$Cd$_{x}$Se/CdS DRs synthesized without precursor-injection delay of the S-precursor. (e) Ensemble photoluminescence spectra and (f) time-resolved PL decays measured at $T\approx 10~\rm{K}$.}
    \label{ensemble_properties}
\end{figure*}

Four Zn$_{1-x}$Cd$_{x}$Se/CdS DR samples (labeled DR\=/1 to DR\=/4) with different $x$ were synthesized  
(for details see Supporting Information). Starting point for samples DR\=/1 to DR\=/3 is the same batch of pre-synthesized ZnSe NCs exhibiting an average diameter of $d=3.16$~nm. A dispersion of these ZnSe NCs as well as a S-precursor solution were separately hot-injected into a Cd-precursor solution at 320\,°C. For sample DR\=/1, ZnSe NCs and the S-precursor were injected simultaneously ($\Delta t=0$ s), while for sample DR\=/2 and DR\=/3 the S-precursor injection was delayed with respect to the ZnSe NCs injection by $\Delta t=10$ s and $\Delta t=60$ s, respectively. The increasing injection delay should increase the degree of cation exchange occurring in the original ZnSe NCs before the CdS-shell growth. After complete shell formation, further cation exchange inside the core is inhibited. Hence, we expect samples DR\=/1 to DR\=/3 to have increasing Cd fractions $x$ in the core. As a reference, sample DR\=/4 represents pure CdSe/CdS DRs (i.\,e.\, $x=1$) that have been synthesized from CdSe NCs (diameter $d=3.18$~nm).

Figure~\ref{ensemble_properties}a--c illustrates results of model calculations within the effective mass approximation showing the theoretical range of band alignments and exciton localizations (see SI). A rising Cd fraction results in a lowered conduction band offset and thus, in a larger electron-hole wave function overlap and a reduced exciton energy. Figure~\ref{ensemble_properties}d shows a representative transmission electron microscopy (TEM) image of sample DR\=/1 (see Figs.\ S1--S3 for TEM images of other samples). The visibly thickened side is known to form around the cores in ZnSe/CdS DRs.\cite{Dorfs.2008,Zhu.2014b} This feature is also present in the CdSe/CdS DRs sample (DR\=/4), possibly resulting from the use of cores with an initial zinc blende phase. Table~\ref{ensemble_properties} gives the average diameters of the DRs at the center ($D$) and at the thickened side ($C$) as well as their average length ($L$). Samples DR\=/1 to DR\=/4 exhibit similar geometries: $D$ and $L$ overlap in their standard deviations. Figure~\ref{ensemble_properties}e depicts normalized ensemble PL spectra of samples DR\=/1 to DR\=/4 at $T \approx 10$~K. The main PL peak red-shifts by about 100~meV in energy from DR\=/1 to DR\=/4,  (Table~\ref{ensemble_properties}). All samples show a broad low-energy band around 1.75~eV with low intensity, indicating trap emission\cite{Thibert.2011,Wen.2012}. The time-resolved PL data in Figure~\ref{ensemble_properties}f reveal decreasing fluorescence lifetimes for the sample series. Fitting the decay curves with biexponential functions yields average PL lifetimes between 172~ns and 34~ns (Table~\ref{ensemble_properties}). Both, the decrease in PL-emission energy and PL lifetime can be explained by an increasing Cd fraction $x$ within the cores, thus, with a transition from type\=/II to type\=/I band alignment.

\begin{table*}[b]
  \caption{Characteristics of the sample series. Core composition, S-injection delay time for CdS shell growth ($\Delta t$), average rod diameter in the center ($D$), at the thickened side ($C$) and averaged DR lengths ($L$), maximum PL emission energies $E_{\rm{PL}}$ of the main peak and intensity averaged fluorescence lifetimes $\tau_{\rm{ave}}$. Geometry parameters were quantified from TEM images from 200 NCs per sample.}
  \label{samples}
  \begin{tabular}{cccccccc}
    \hline
    Sample & Core & $\Delta t$ (s) & $D$ (nm) & $C$ (nm) & $L$ (nm) & $E_{\rm{PL}}$ (eV) 
    &$\tau_{\rm{ave}}$ (ns)\\
    \hline
    DR\=/1&ZnCdSe & 0 & 4.8 $\pm$ 0.7 &7.2 $\pm$ 1.7& 27.4 $\pm$ 2.9& 2.120 & 172\\
    DR\=/2&ZnCdSe & 10 & 5.2 $\pm$ 0.7 &6.2 $\pm$ 1.0& 28.8 $\pm$ 3.0 & 2.094 & 126\\
    DR\=/3&ZnCdSe & 60 & 4.8 $\pm$ 0.6 &6.0 $\pm$ 1.3& 25.2 $\pm$ 3.6 & 2.051 & 67\\
    DR\=/4&CdSe & 0 & 4.3 $\pm$ 0.5&5.4 $\pm$ 1.5& 21.7 $\pm$ 11.6& 2.031 & 34\\
    \hline
  \end{tabular}
\end{table*}

Determining the exact material composition of the cores is difficult. In an earlier work we have estimated the Cd fraction for sample DR\=/1 to be $x \approx 50\%$\cite{Rebmann.2023}. Additionally, we quantified the Cd fraction via extended X-ray absorption fine structure (EXAFS) spectroscopy (see SI). For these measurements, an additional sample set DR\=/1' to DR\=/3' was synthesized under the same conditions with slightly smaller ZnSe NCs (diameter of 2.86~nm instead of 3.16~nm). Evaluation of the EXAFS data (see SI) yields values of $x= 0.63$, 0.73, and 0.77 for S-injection delay times of $\Delta t=0$~s, 10~s, and 60~s, respectively. Due to the smaller diameter, these values represent an upper limit of the values expected in samples DR\=/1 to DR\=/3.

\begin{figure}[t]
    \includegraphics[width= 5cm]{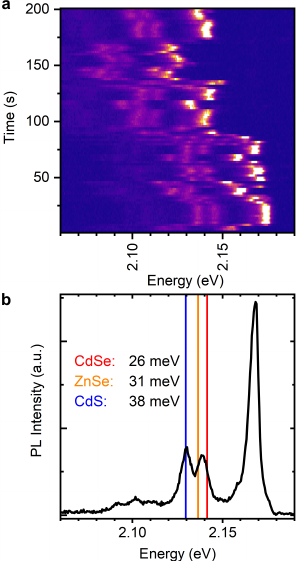}
    \caption{Single-particle spectroscopy at $T\approx 10 \rm{K}$. (a) Series of subsequently measured PL spectra of a single NC from sample DR\=/2, each collected with an integration time of 2~s. (b) PL spectrum of the same DR created by averaging multiple spectra with a similar ZPL energy. The energies of the first phonon lines are highlighted based on bulk values\cite{Madelung.2004}. The particle was excited at 2.78~eV with 10 Mhz repetition rate.
    }
    \label{low_temp}
\end{figure}

In the following, we discuss single-nanocrystal PL spectra measured at $T\approx 10~\rm{K}$. Figure~\ref{low_temp}a shows the temporal evolution of the PL spectrum of an exemplary single Zn$_{1-x}$Cd$_{x}$Se/CdS DR from sample DR\=/2. The brightest peak corresponds to the zero-phonon line (ZPL), while the energetically lower peaks are phonon replica arising from exciton--phonon coupling. All peaks exhibit collective jittering and spectral jumps in time. This spectral diffusion originates from the formation, movement, and disappearance of surface charges.\cite{Empedocles.1996,Fernee.2010} Spectral diffusion is especially pronounced for a type\=/II band alignment, since here one of the charge carriers is mainly located in the shell, making the exciton more sensitive to the impact of surface charges. 

Single-particle spectral time traces were analyzed by averaging those individual spectra within a time trace that exhibit a similar ZPL energy (in the range of 1~nm, i.e., $\Delta E\approx 2$~meV). For the spectral time trace shown in Figure~\ref{low_temp}a, this results in a spectrum with increased signal-to-noise ratio as depicted in Figure~\ref{low_temp}b. Here, the ZPL occurs at $E_\mathrm{ZPL} = 2.168$~eV. Two LO-phonon replica can be distinguished as first order phonon lines separated from the ZPL by 28~meV and 38~meV, respectively. Signatures of higher-order replica are visible in a broader range around 70 meV below the ZPL energy. 
For comparison, literature values of the bulk LO phonon energies\cite{Madelung.2004} of CdSe, ZnSe, and CdS are plotted as vertical lines relative to $E_\mathrm{ZPL}$. Since the first LO-phonon replica occurs between the references of CdSe and ZnSe, it is called a Se-type replica, with an energy $\Delta E_\mathrm{Se}$ below $E_\mathrm{ZPL}$. The second LO-phonon replica, which matches the CdS-phonon reference, is called S-type replica. 

\begin{figure*}[t]
    \includegraphics[width=160mm]{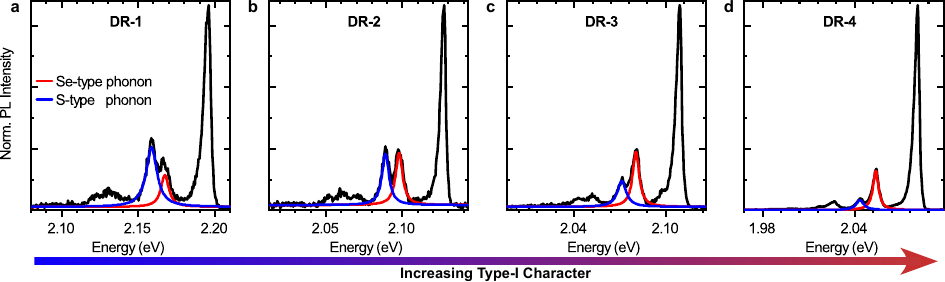}
    \caption{Representative averaged PL spectra of single DRs from (a) sample DR-1 to (d) sample DR-4 measured at $T\approx 10 \rm{K}$. The spectra were normalized to their ZPL intensity. The respective first order phonon peaks are highlighted with Lorentzian fits.}
    \label{phonon_trend}
\end{figure*}

Averaging of the spectra for one ZPL per NC was performed on approximately 100 DRs for each sample. From these particles, 40--70 \% provided spectra with a sufficient signal-to-noise ratio concerning the LO-phonon replica. The average Se-type phonon energies are $\Delta E_\mathrm{Se}$ 28~$\pm$~0.7~meV (DR\=/1), 28~$\pm$~0.5~meV (DR\=/2), 28~$\pm$~0.7~meV (DR\=/3), and 27~$\pm$~0.4~meV (DR\=/4). These values do not indicate a conclusive trend. For a gradual increase in Cd fraction $x$ in the Zn$_{1-x}$Cd$_{x}$Se core, one might expect a continuous decrease of $\Delta E_\mathrm{Se}$.\cite{Azhniuk.2009} If existing, this trend seems to be too small to be resolved with respect to the emission-line widths of the presented data. However, consistent with above considerations, DR\=/4 shows the lowest value of $\Delta E_\mathrm{Se}$.
The average S-type phonon energies $\Delta E_\mathrm{S}$ are 36~$\pm$~0.4~meV (DR\=/1), 38~$\pm$~0.3~meV (DR\=/2), 37~$\pm$~0.7~meV (DR\=/3) and 36~$\pm$~0.6~meV (DR\=/4). Here, no significant trend can be deduced, nor would this have been expected. 

Even though the energy of the phonon replica cannot be correlated to the Cd fraction $x$, a correlation of their relative intensities is possible, as will be shown below for the first-order replica. Figure~\ref{phonon_trend} displays representative averaged PL spectra of individual particles for each sample, normalized to the ZPL-intensity maximum. These spectra suggest that the coupling to LO phonons generally decreases with increasing Cd content in the core, corresponding to an increase in the type\=/I character. The relative intensity of the first order S-type phonon decreases compared to the Se-type phonon across the sample series.

\begin{figure*}[t]
    \includegraphics[width=166.947mm]{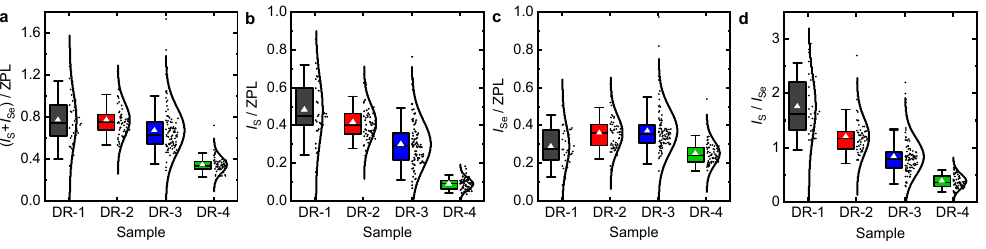}
    \caption{
    Statistical evaluation of the first order phonon intensities from single DRs. (a--c) Ratios of the sum of (a) S- and Se-type, (b) S-type, and (c) Se-type phonon intensity with respect to the ZPL intensity. (d) Ratio of the S- and Se-type phonon intensities. The boxes cover the interquartile range, while the whiskers show the 95th percentile. The white triangles indicate the average values and the horizontal lines inside the boxes indicate the median values. On the right side of the boxes each black dot corresponds to the phonon ratio measured from a single DR. In the respective samples DR-1 to DR-4 33, 59, 91 and 63 single DRs were evaluated. The curves fit the data points based on normal distributions.
    }
    \label{statistics}
\end{figure*}

To quantify the phonon coupling, the peaks in each averaged PL spectrum of individual DRs were fitted by Lorentzian functions and the following combinations of first order and ZPL intensity ratios were evaluated: ($I_\mathrm{S}+I_\mathrm{Se})/I_\mathrm{ZPL}$, $I_\mathrm{S}/I_\mathrm{ZPL}$, $I_\mathrm{Se}/I_\mathrm{ZPL}$, and $I_\mathrm{S}/I_\mathrm{Se}$. Figure~\ref{statistics} summarises their statistical distributions (for further statistics see Table\ S2). The intensities are not uniform and scatter to different degrees. The relative total first-order LO-phonon coupling ($I_\mathrm{S}+I_\mathrm{Se})/I_\mathrm{ZPL}$ (Fig.\ \ref{statistics}a) decreases within the sample series, i.e.\ with increasing type\=/I character. The relative S-type phonon intensity $I_\mathrm{S}/I_\mathrm{ZPL}$ (Fig.\ \ref{statistics}b) gradually decreases with increasing type\=/I character, while the Se-type phonon intensity $I_\mathrm{Se}/I_\mathrm{ZPL}$ (Fig.\ \ref{statistics}c) follows no apparent trend and changes on a smaller scale. The intensity ratio of the S-type to Se-type phonon $I_\mathrm{S}/I_\mathrm{Se}$ (Fig.\ \ref{statistics}d), which describes the coupling independent of the ZPL, decreases with increasing type\=/I character.

In the following, we qualitatively discuss the behavior of the phonon-replica intensities. In ionic semiconductors the exciton-LO-phonon coupling is primarily facilitated by the Fröhlich interaction\cite{Frohlich.1954}, which describes the coupling of the excitonic charge distribution to the LO phonon via Coulomb interaction.\cite{Gindele.1999,Kalliakos.2002} The decrease of the total LO-phonon coupling apparent in Fig.\ \ref{statistics}a can be explained with a larger charge-carrier overlap for the type\=/I band alignment to the type\=/II case.\cite{Nomura.1992, Empedocles.1999b} The major change of the S-type phonon (Fig.\ \ref{statistics}b) and the minor change of the Se-type phonon intensity (Fig.\ \ref{statistics}c) correlates well with changes of the charge-carrier localization within the sulfide shell and the selenide core of the DRs (\ref{ensemble_properties}a--c). As the band alignment changes from type\=/II to type\=/I the phonon coupling to the S-type mode is decreased. The broad spread of coupling strengths within each sample (Fig.\ \ref{statistics}a-d) can be a result of the particle-size distribution, different degrees of cation exchange, and varying surface-charge constellations. For reference, Empedocles et al.\cite{Empedocles.1996} found strong deviations of exciton--phonon coupling for single CdSe and CdSe/ZnS spherical NCs, between 0.06--1.3. Deviations have also been observed for nominally identical vapor-phase grown InGaAs NCs.\cite{Dufaker.2010} 

A quantitative modeling of the coupling of excitons to the different LO-phonon modes in heterostructured DRs is a complicated task. Only recently, Lin et al.\cite{Lin.2023} modeled the coupling for spherical type-I CdSe/CdS NCs atomistically. A quantitative modeling of anisotropic type-II and type-I heterostructured DRs as investigated here has not been accomplished, yet, and is beyond the scope of this work. 

The exciton--phonon coupling strength is commonly described by the Huang--Rhys factor\cite{Huang.1950}, which is given by the intensity ratio of the first order phonon-replica to the ZPL. In heteronanostructures, in particular with a type-II band alignment, this approach has to be modified to account for different couplings in different materials of both electrons and holes. It is well known that electrons and holes couple differently to phonons.\cite{Devreese.2015} We propose a simple model in which the coupling is proportional to the probability\cite{Nomura.1992,Dufaker.2010,Callsen.2015,Iaru.2021} of the charges in the shell or the core. For the relative S-type and Se-type phonon replica intensities, we presume

\begin{equation}\label{ratio1}
\frac{ I_\mathrm{S} }{I_\mathrm{ZPL}} = 
\left|\beta_\mathrm{S}^{\rm{e}} \cdot \rho_\mathrm{shell} ^{\rm{e}}  -
\beta_\mathrm{S}^{\rm{h}} \cdot  \rho_\mathrm{shell} ^{\rm{h}} 
\right|
\end{equation}
and 
\begin{equation}\label{ratio2}
\frac{ I_\mathrm{Se} }{I_\mathrm{ZPL}} = 
\left|\beta_\mathrm{Se}^{\rm{e}} \cdot \rho_\mathrm{core} ^{\rm{e}}  -
\beta_\mathrm{Se}^{\rm{h}} \cdot  \rho_\mathrm{core} ^{\rm{h}} 
\right|,
\end{equation}
respectively, with coupling constants $\beta_{\mathrm{S,Se}}^{\rm{{e,h}}}$ and probabilities 
\begin{equation}
\rho_\mathrm{shell,core}^{\rm{{e,h}}}=\int_{\rm{shell,core}} |\psi_{\mathrm{{e,h}}} |^2  \rm{d}V.
\end{equation}
This approach accounts for the possibility that overlapping electron and hole wavefunctions $\psi_{\rm{e,h}}$ can suppress the overall phonon coupling. 

\begin{figure*}[t]
    \includegraphics[width=167.725 mm]{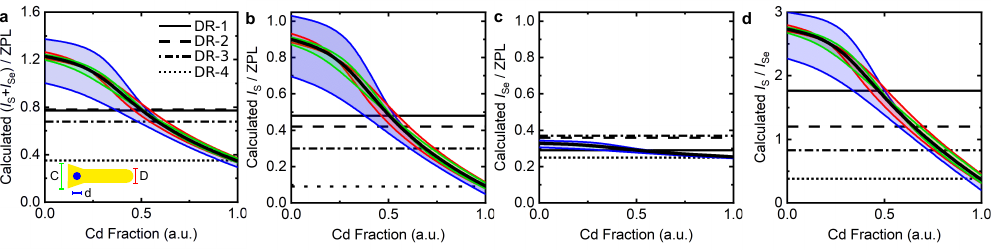}
    \caption{Estimation of the properties depicted in Figure~\ref{statistics} based on theoretical calculations within the effective mass approximation. The following phonon ratios are illustrated: (a) $(I_\mathrm{S}+I_\mathrm{Se})/I_\mathrm{ZPL}$, (b) $I_\mathrm{S}/I_\mathrm{ZPL}$, (c) $I_\mathrm{Se}/I_\mathrm{ZPL}$ and (d) $I_\mathrm{S}/I_\mathrm{Se}$.
    The black curves were calculated with the average geometry of the samples. The colored lines were calculated by independently changing the core ($d=3.17\pm0.43$~nm), rod ($D=4.78\pm0.62$~nm) and cone ($C=6.21\pm1.40$~nm) diameters according to their averaged standard deviations obtained from TEM data and keeping the rod length at 25.78~nm. The inset depicts the schematic cross-section that was used in the calculations. The horizontal lines indicate the average experimental values.
    }
    \label{coupling}
\end{figure*}

The probabilities were calculated within the effective mass approximation\cite{Panfil.2022,Park.2012} (Details see SI). The geometry of the DRs was approximated by a spherical core (diameter $d$), embedded in a rod-shaped shell (diameter $D$, length $L$). The core was placed on one end of the rod such that it was covered by 3 monolayers of shell material.  
The bulge-like shape of the DRs is modeled as a truncated cone (base diameter $C$, height $C/2$, cutting-surface diameter $D$; see inset of Fig.~\ref{coupling}). Average geometry parameters that represent samples DR\=/1 to DR\=/4 were determined from spectroscopy of the core sample ($d$) and from TEM investigations ($D,L$, and $C$). For the Zn$_{1-x}$Cd$_x$Se-core material, a homogeneous cation distribution was assumed.\cite{Zhong.2007,Groeneveld.2013,Boldt.2020,Rebmann.2023} 

As mentioned above, the Cd content $x$ in the Zn$_{1-x}$Cd$_x$Se-core of samples DR\=/1 to DR\=/3 is subject to some uncertainties. It is best known for sample DR\=/1, for which we assume $x=0.50$.\cite{Rebmann.2023} For sample DR\=/4, $x=1 $ is valid. With that, we calculated 
$\rho_\mathrm{shell,core}^{\rm{e,h}}$ for samples DR\=/1 and DR\=/4. Together with the experimentally determined values of the ratios $I_\mathrm{S,Se}/I_\mathrm{ZPL}$ for both samples, equations \ref{ratio1} and \ref{ratio2} represent a linear system of equations with the four unknowns $\beta_{\mathrm{S,Se}}^{\rm{e,h}}$. Solving this system of equations yields the values $\beta_{\mathrm{S}}^{\rm{e}}=1.35$, $\beta_{\mathrm{S}}^{\rm{h}}=3.35$, $\beta_{\mathrm{Se}}^{\rm{e}}=0.12$, and $\beta_{\mathrm{Se}}^{{\rm{h}}}=0.38$. 
Interestingly, these values exhibit similarities to the bulk Fröhlich-coupling constants $\alpha$ (see Table~S4). Like the corresponding Fröhlich constants, the coupling constants $\beta$ are larger for the hole than for the electron and they are larger for CdS than for ZnSe and CdSe. Furthermore, the ratio $\beta_{\mathrm{S}}^{\rm{e}}/\beta_{\mathrm{S}}^{\rm{h}}$ is similar to $\alpha_{\mathrm{S}}^{\rm{e}}/\alpha_{\mathrm{S}}^{\rm{h}}$.

The coupling constants $\beta_{\mathrm{S,Se}}^{\rm{{e,h}}}$ have then been used --- together with results from COMSOL simulations --- to estimate the relative phonon-coupling strengths. Results are given in Figures \ref{coupling}, where panel b and c represent relative S- and Se-type phonon intensities calculated by equations \ref{ratio1} and \ref{ratio2}, respectively, while the consequential ratios $(I_\mathrm{S}+I_\mathrm{Se})/I_\mathrm{ZPL}$ and $I_\mathrm{S}/I_\mathrm{Se}$ are depicted in panels a and d, respectively, analogous to Fig.\ \ref{statistics}.

In each panel, the black curve is obtained for a geometry averaged over all four samples.  The horizontal lines represent the experimental average values for DR\=/1--4. The modeling demonstrates that the coupling strength to the S-type phonon is strongly decreasing with increasing $x$ (see Figure~\ref{coupling}b). This is in line with the experimentally observed decrease from about 0.5 for $x=0.5$ (DR\=/1) to 0.1 for $x=1$ (DR\=/4). The modeling results estimate the Cd fraction of samples DR\=/2 and DR\=/3 to 0.57 and 0.69, respectively.
The modeled coupling to the Se-type phonon is essentially constant (between 0.35 and 0.25 for $0<x<1$, see Figure~\ref{coupling}c). A subtle increase in coupling for DR\=/2 and DR\=/3 observed in the experiment cannot be reproduced by the model. The modeled total coupling strength in Figure~\ref{coupling}a decreases with increasing $x$; it is dominated by the coupling to the S-type phonon. 
Figure~\ref{coupling}d shows the coupling ratio of the S- to Se-type phonon intensity, independently of the ZPL. Using this data to again estimate $x$ for samples DR\=/2 and DR\=/3 yields values around 0.65 and 0.79, respectively. These values are on a similar scale as the values determined by EXAFS spectroscopy. 

Numerical modeling allows to investigate the influence of each geometry parameter on the intensity ratios. For this the calculations were repeated and the diameters of the rod $D$, cone-base $C$, or core $d$ were separately increased or decreased by their experimental standard deviation. Results are given by the colored curves in Fig.\ \ref{coupling}.
It can be deduced that the spread of different core diameters within a sample has the strongest effect, while the cone and rod diameters have a weaker effect. The position of the core and different band offsets have a neglectable influence (see Fig.\ S8).
Considering the variation of all geometry parameters together, one would expect a broader distribution of phonon-coupling ratios for small $x$. This agrees with the experimentally observed distributions of data points in Fig.\ \ref{statistics}.

A closer look on the data of Fig.\ \ref{statistics}d reveals that the standard deviation for sample DR\=/1 is a factor of 4 larger than for DR\=/4. From the calculated data, summing the deviations for all geometric variations, one would expect the broadening for $x=0.5$ to be only a factor of 1.5 larger than for $x=1$. Therefore an additional contribution might be relevant. It has been reported that surface charges and point defects strongly increases the exciton--phonon coupling in NCs, compared to the intrinsic coupling.\cite{Nomura.1992,Krauss.1997,Sagar.2008,Cui.2016,Iaru.2021} In order to estimate the influence of surface charges in our system, we placed a positive or negative point charge on the model structure at the position where the conical shape changes into the cylindrical shape (cf.\ inset of Fig.\ \ref{coupling}a). The calculated data shown in Fig.\ S9 reveal a strong change of the S-type phonon coupling, which is maximal for $x\approx 0.5$.  
Thus, surface charges can explain the experimentally observed broader distribution of S- to Se-type phonon ratios for DR\=/1. 
In particular negative surface charges strongly increase the S-type phonon coupling for $x$ around 0.5. These charges might cause the experimentally determined averages of the relative S-type phonon intensities to be larger than their median values, as depicted in Figure~\ref{statistics}b and d.

In conclusion, we investigated the charge-carrier localization in Zn$_{1-x}$Cd$_{x}$Se/CdS DRs with different compositions. Optical and X-ray spectroscopy proved that in situ cation exchange is suitable for band engineering. This allowed us to design a set of samples with different band alignments, while keeping the geometry comparable. Statistical data obtained from cryogenic-temperature spectroscopy on the single-particle level revealed that the first-order LO-phonon ratio of the S- to Se-type phonon replica changes within the set of samples and is thus an indicator of different charge-carrier localization. Our simplified theoretical model for the coupling ratios captures the general trend of the phonon ratios. Our study suggests that exciton--phonon coupling could serve as a new quantity characterizing the exciton localization. For a better understanding of exciton--phonon coupling, atomistic calculations are still being actively developed.\cite{Lin.2023,Han.2019,Han.2022,Jasrasaria.2021,Lin.2023}

\paragraph{}
\textbf{Notes}\\
The authors declare no competing financial interest.
\\
\begin{acknowledgement}
The electron microscopy service of the University of Hamburg is greatfully acknowledged for TEM measurements. We acknowledge the DESY (Hamburg, Germany), a member of the Helmholtz Association (HGF), for the provision of experimental facilities. Parts of this research were carried out at Petra III and we would like to thank Dr.\ Wolfgang Caliebe for assistance in using Beamline P64. 

This work was funded by the Deutsche Forschungsgemeinschaft (DFG) via GRK 2536 NANOHYBRID, project number 406076438 and via the ‘CUI: Advanced Imaging of Matter’ EXC 2056, project number 390715994. D.K., L.K. and J.K. acknowledge funding from the ERC Consolidator Grant LINCHPIN (grant no. 818941).
\end{acknowledgement}

\begin{suppinfo}

The following files are available free of charge.
\begin{itemize}
  \item Supporting Information: Description of the synthesis, TEM data, UV/Vis spectra, EXAFS analysis, single-particle statistics from PL measurements and calculation details.
\end{itemize}

\end{suppinfo}

\bibliography{literature.bib}
\newpage
\begin{figure*}[]
    \includegraphics[width= 8.25 cm]{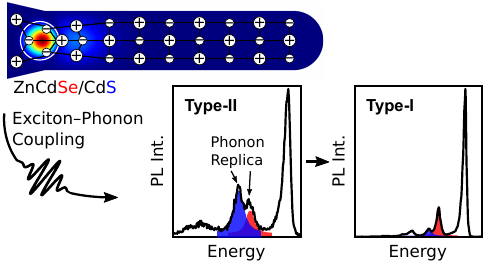}
    \caption{For Table of Contents Only}
    \label{ToC}
\end{figure*}
\end{document}